\documentclass{iopart}
\usepackage{iopams}
\usepackage{graphicx}
\begin{document}

\title[Phase diagrams of correlated electrons: systematic corrections]
{Phase diagrams of correlated electrons: systematic corrections
to the mean field theory}

\author{R. Hlubina$^{1,2}$}

\address{$^1$Division of Solid State Physics, Comenius University,
Mlynsk\'{a} Dolina F2, 842 48 Bratislava, Slovakia}

\address{$^2$International School for Advanced Studies (SISSA), Via Beirut
2-4, I-34014 Trieste, Italy}

\begin{abstract}
Perturbative corrections to the mean field theory for particle-hole
instabilities of interacting electron systems are computed within a scheme
which is equivalent to the recently developed variational approach to the
Kohn-Luttinger superconductivity.  This enables an unbiased comparison of
particle-particle and particle-hole instabilities within the same
approximation scheme. A spin-rotation invariant formulation for the
particle-hole instabilities in the triplet channel is developed. The method is
applied to the phase diagram of the $t$-$t^\prime$ Hubbard model on the square
lattice. At the Van Hove density, antiferromagnetic and $d$-wave Pomeranchuk
phases are found to be stable close to half filling. However, the latter phase
is confined to an extremely narrow interval of densities and away from the
singular filling, $d$-wave superconducting instability dominates.
\end{abstract}

\section{Introduction}
In strongly correlated electron systems, usually several symmetry breaking
states appear as natural candidates for the ground state. This is so because,
due to the large interaction strength, typically various mean field criteria
for an instability are simultaneously satisfied. Unfortunately, mean field
theory can not tell reliably which of the possible competing instabilities
wins. Therefore we are forced to look for finer methods, taking into account
also the correlations which are neglected at the mean field level. The present
paper is concerned with developing a systematic method that includes the
dominant correlation effects neglected in the mean field theory. We require
that the method satisfies the following criteria.

(i) As emphasized by Anderson \cite{Anderson84}, the energy of a quantum state
is dominantly determined by its short-range correlations. Therefore, an
unbiased comparison of different symmetry breaking patterns requires that the
same 'amount' of correlations is kept in all competing states. Obviously, mean
field theory satisfies this criterion, neglecting the correlations in all
states. 

(ii) Long ago it has been shown that, when treated beyond the mean field
level, even purely repulsive systems can support superconductivity
\cite{Kohn65}. We require that the general method reduces, in the case of
superconductivity, to the recently developed perturbative scheme which enables
a variational treatment of the Kohn-Luttinger superconductors \cite{Mraz03}.

Combining the above requirements, we are led to the search for a general
lowest-order perturbative correction scheme to the mean field theory which
treats all symmetry breaking states on the same footing.  This goal will be
accomplished as follows. First we choose the type of instability we want to
study. Then we split the Hamiltonian $H$ of the interacting system into two
parts. The first part contains the kinetic energy $H_0=\sum_{{\bf
k},\sigma}\varepsilon_{\bf k}n_{{\bf k},\sigma}$ and that part $H_1$ of the
interaction Hamiltonian which corresponds to a generalization of the reduced
interaction Hamiltonian of the BCS theory \cite{Bardeen57}. For instance, in
the case of superconductivity, $H_1$ describes scattering of the Cooper pairs
with zero total momentum, or in the case of magnetism, scattering of triplet
particle-hole pairs with a given total momentum ${\bf q}$. Denoting the rest
of the interaction Hamiltonian as $H_2$, the total Hamiltonian can be written
as $H=H_0+H_1+H_2$.

As shown in the case of superconductivity by Bogoliubov \etal
\cite{Bogoliubov61}, the mean field approach, when applied to the Hamiltonian
$H_0+H_1$, is essentially exact in the thermodynamic limit.  In order to
preserve this exactness, the term $H_2$, which describes scattering processes
not included in the mean field theory, should be eliminated. Within our
approach, this goal is achieved perturbatively. Namely, we construct a
canonical transformation $\tilde{H}=e^{iS}He^{-iS}$ which eliminates
scattering processes to first order in the interaction strength.  This is
achieved for $H_2+i[S,H_0]=0$ and the effective Hamiltonian reads, to second
order in the interaction strength, as $\tilde{H}=H_0+H_1+\tilde{H}_2$, where
$\tilde{H}_2=[iS,H_1+H_2/2]$.  The term $\tilde{H}_2$, which has replaced the
term $H_2$ in the original Hamiltonian $H$, is typically even more complicated
than $H_2$. So what have we gained?  The key observation of our approach is
that at weak coupling, $\tilde{H}_2$, being of second order in the coupling
constant, is much smaller than $H_2$ and therefore can be treated in the mean
field approximation.  Note that this yields a nontrivial correction to the
original mean field theory, because $\tilde{H}_2$ contains in general also
terms of the type singled out into $H_1$.  In order to better understand the
last point, it is useful to observe that our approach is essentially
variational, with a mean-field type ansatz for the wavefunction
$|{\tilde\psi}\rangle$ which minimizes the ground state expectation value of
the energy,
$E=\langle{\tilde\psi}|\tilde{H}|{\tilde\psi}\rangle$. Equivalently, one can
write $E=\langle\psi|H|\psi\rangle$ with
$|\psi\rangle=e^{-iS}|{\tilde\psi}\rangle$ and interpret $|\psi\rangle$ as a
mean field state written in terms of quasiparticles. In this language it is
not surprising that the effective interaction between the quasiparticles
differs from the bare interaction.

The plan of the paper is as follows. In Section 2 we construct the canonical
transformation for the Hubbard model. We show explicitly that the effective
Hamiltonian $\tilde{H}$ depends on the instability channel. It is worth
pointing out that this is a common feature of our method and of various
versions of the renormalization group approach to correlated
electrons \cite{Shankar94,Zanchi97,Wegner01,Honerkamp01,Hankevych03}.

In Section 3 we show how the method introduced in \cite{Mraz03} fits the
general scheme and we briefly review the main results of the method when
applied to the superconducting instability.

In Section 4 we develop the formalism for particle-hole instabilities with a
finite total momentum of the particle-hole pairs. Our formulation allows for a
simultaneous discussion of both the singlet and the triplet channels, the
triplet channel being analyzed within an explicitly spin-rotation invariant
formalism. 

In Section 5 we study particle-hole instabilities with a vanishing total
momentum of the particle-hole pairs, called Pomeranchuk instabilities within
Landau's Fermi liquid theory \cite{Pomeranchuk58}. We take into account
simultaneously the singlet and the triplet sectors, and also all point group
symmetries, except for the identical representation in the singlet sector. For
lattice models, this latter 'instability' is not associated with any symmetry
breaking and simply corresponds to a change of the electron dispersion.

Finally, in Section 6 we present the results of explicit calculations for the
$t$-$t^\prime$ Hubbard model on the square lattice at electron densities
$\rho$ corresponding to the Van Hove filling of the noninteracting band, and we
specialize to $\rho\sim 1$ and small $t^\prime/t$.

\section{Canonical transformation}
For the sake of simplicity, let us proceed with an explicit evaluation
of $\tilde{H}$ for the minimal microscopic model of interacting electrons,
namely the Hubbard model. In that case we can write
\begin{eqnarray*}
H_1&=&{U\over L}\sum_{\{123\}}
c^\dagger_{3 \uparrow} c_{1 \uparrow}
c^\dagger_{4 \downarrow} c_{2 \downarrow}\Delta_{1234},
\\
H_2&=&{U\over L}\sum_{\{123\}}
c^\dagger_{3 \uparrow} c_{1 \uparrow}
c^\dagger_{4 \downarrow} c_{2 \downarrow}
\left(1-\Delta_{1234}\right),
\end{eqnarray*}
where $L$ is the number of lattice sites and the summation index $\{123\}$
means momentum conservation ${\bf k}_1+{\bf k}_2={\bf k}_3+{\bf k}_4$. The
cutoff function $\Delta_{1234}$ is always zero, except for the scattering
processes which are to be singled out into $H_1$ in which case
$\Delta_{1234}=1$. For technical reasons, we also take $\Delta_{1234}=1$ for
processes which conserve the energy, i.e. if $\varepsilon_{1}+\varepsilon_{2}
=\varepsilon_{3}+\varepsilon_{4}$. This latter assumption will always be made
in this paper and, for the sake of simplicity, it will not be mentioned
explicitly in the formulas that follow.

One verifies readily that the canonical transformation we look for
has the following Hermitian generator,
\begin{equation}
S={iU\over L}\sum_{\{123\}}
{{1-\Delta_{1234}}\over
{\varepsilon_{1}+\varepsilon_{2}-\varepsilon_{3}-\varepsilon_{4}}}
c^\dagger_{3 \uparrow} c_{1 \uparrow}
c^\dagger_{4 \downarrow} c_{2 \downarrow}.
\end{equation} 
It is worth pointing out that for 
\begin{equation}
\Delta_{1234}=\Delta_{2134}=\Delta_{1243},
\label{eq:spin_rotation_invariance}
\end{equation}
which will be always assumed from now on, the generator is explicitly spin
rotation-invariant. A straightforward calculation shows that the effective
scattering term $\tilde{H_2}$ is
\begin{eqnarray}
\fl
&\tilde{H_2}=-{U^2\over 2L^2}\sum_{\{123\}}
\sum_{\{\alpha\beta\gamma\}}
{{(1-\Delta_{1234})(1+\Delta_{\alpha\beta\gamma\delta})}\over
{\varepsilon_{1}+\varepsilon_{2}-\varepsilon_{3}-\varepsilon_{4}}}
\nonumber
\\
\fl
&\times\left[(\delta_{1\gamma}c^\dagger_{3\uparrow} c_{\alpha\uparrow}
-\delta_{3\alpha}c^\dagger_{\gamma\uparrow} c_{1\uparrow})
c^\dagger_{4\downarrow} c_{2\downarrow}
c^\dagger_{\delta\downarrow} c_{\beta\downarrow}
+(\delta_{2\delta}c^\dagger_{4\downarrow} c_{\beta\downarrow}
-\delta_{4\beta}c^\dagger_{\delta\downarrow} c_{2\downarrow})
c^\dagger_{\gamma\uparrow} c_{\alpha\uparrow}
c^\dagger_{3\uparrow} c_{1\uparrow}\right].
\label{eq:H_eff}
\end{eqnarray}
Note that $\tilde{H_2}$ describes three-particle collisions and as such
(together with terms generated from higher orders of perturbation theory)
allows for more complicated patterns of symmetry breaking than simple
particle-particle or particle-hole condensates.  Here we do not consider such
possibilities.

In order to proceed it is useful to specialize at this point to a particular
symmetry breaking channel. In this paper we will discuss three channels: (i)
superconductivity with Cooper pairs with total momentum ${\bf q}=0$, (ii)
particle-hole pairs with a finite total momentum ${\bf Q}$, and (iii)
particle-hole pairs with a vanishing total momentum ${\bf q}=0$. It is
possible to apply our method also to superconductivity with Cooper pairs with
a nonvanishing total momentum, but we have not done so.  As will become clear
later, in each symmetry breaking channel, there are singlet and triplet
sectors in the spin space and various representations of the point group,
therefore the number of different symmetry breaking states is enormous.

\section{Superconducting channel}
This case has been discussed in detail in
\cite{Mraz03,Mraz04a,Mraz04b,Mraz05}, so the discussion will be very brief.
According to the BCS theory \cite{Bardeen57}, superconductivity can be viewed
as an instability of the symmetric phase with respect to formation of Cooper
pairs with zero total momentum. Therefore in the superconducting channel we
choose
\begin{equation*}
\Delta_{1234}=\cases{1&for ${\bf k}_1+{\bf k}_2={\bf k}_3+{\bf k}_4=0$\\
0&otherwise\\}
\end{equation*}
Taking the expectation value of $\tilde{H}$ in the superconducting state and
introducing the angle-resolved pair field $b_{\alpha\beta}({\bf k})=\langle
c_{-{\bf k}\alpha}c_{{\bf k}\beta}\rangle$, we have
\begin{eqnarray}
E=\sum_{{\bf k}\sigma}\varepsilon_{\bf k}f_{{\bf k}\sigma}
-{1\over 2L}\sum_{{\bf k},{\bf p}} V_{{\bf k} {\bf p}}^{\rm sc}
b_{\alpha\beta}^\ast({\bf k})b_{\alpha\beta}({\bf p})
+E_{\rm FL},
\label{eq:sc_energy}
\end{eqnarray}
where 
\begin{equation*}
E_{\rm FL}={U\over L}{N^2\over 4}
+{U^2\over L^2}\sum_{\{123\}}
{{f_{1}f_{2}(1-f_{3})(1-f_{4})}
\over
{\varepsilon_{1}+\varepsilon_{2}
-\varepsilon_{3}-\varepsilon_{4}}}
\end{equation*}
is that part of the interaction energy which is not associated with symmetry
breaking.  Note that we have chosen an opposite sign convention for $V_{{\bf
k} {\bf p}}^{\rm sc}$ with respect to \cite{Mraz03,Mraz04a,Mraz04b,Mraz05}.
In the present paper attractive interactions are chosen to be positive. For
exponentially small order parameters, which we expect in the validity region
of the present theory, we can neglect the change of $E_{\rm FL}$ between the
symmetric and symmetry broken state and this term will not be discussed any
more.

The pair scattering amplitude reads
\begin{equation*}
V_{{\bf k} {\bf p}}^{\rm sc}=-U-U^2 
\chi_{\rm ph}^\prime({\bf k}+{\bf p},\varepsilon_{\bf p}-\varepsilon_{\bf k}),
\end{equation*} 
where $\chi_{\rm ph}^\prime({\bf q},\omega)$ is the real part of the
particle-hole susceptibility 
\begin{equation*}
\chi_{\rm ph}({\bf q},\omega)={1\over L}\sum_{\bf K}
{f_{\bf K}-f_{{\bf K}+{\bf q}}\over \varepsilon_{{\bf K}+{\bf
q}}-\varepsilon_{\bf K}-\omega-i0}.
\end{equation*}
The imaginary part of $\chi_{\rm ph}({\bf q},\omega)$ does not contribute to
$V_{{\bf k} {\bf p}}^{\rm sc}$, because the energy conserving processes are
excluded from $H_2$. In the actual numerical implementation we neglect the
feedback effects on the susceptibility and replace the occupation numbers
$f_{\bf k}$ by their values in the noninteracting system, $f^0_{\bf k}$. This
replacement is again well controlled for exponentially small order parameters.

Following \cite{Balian63,Anderson75} we introduce a $2\times 2$ matrix
notation $\hat{b}_{\bf k}$ for the pair field with the matrix elements
$(\hat{b}_{\bf k})_{\alpha\beta}=b_{\alpha\beta}({\bf k})$. Anticommutation of
fermion operators then implies $\hat{b}_{-\bf k}=-\hat{b}_{\bf k}^T$ where
$X^T$ is a matrix transposed to $X$. Let us further define the gap matrix
$\hat{\Delta}_{\bf k}={1\over L}\sum_{\bf p}V_{{\bf k}{\bf p}}\hat{b}_{\bf p}$
with the same symmetry properies as $\hat{b}_{\bf k}$ and introduce the
following parameterization
\begin{equation*}
\hat{\Delta}_{\bf k}=
\left(\begin{array}{cc}
-d^1_{\bf k}+id^2_{\bf k},&d^3_{\bf k}+d^0_{\bf k}\\
d^3_{\bf k}-d^0_{\bf k},&d^1_{\bf k}+id^2_{\bf k}
\end{array}\right),
\end{equation*}
where a complex four-vector field $d^\nu_{\bf k}=(d^0_{\bf k},{\bf d}_{\bf
k})$ has been introduced which satisfies the conditions $d^0_{-\bf k}=d^0_{\bf
k}$ and ${\bf d}_{-\bf k}=-{\bf d}_{\bf k}$. 

States for which the vector function ${\bf q}_{\bf k}= d^0_{\bf k}{\bf d}_{\bf
k}^\ast+(d^0_{\bf k})^\ast{\bf d}_{\bf k} +i{\bf d}_{\bf k}\times {\bf d}_{\bf
k}^\ast$ vanishes identically are called unitary \cite{Mraz05}, since in that
case $\hat{\Delta}_{\bf k}\hat{\Delta}_{\bf k}^\dagger$ is proportional to a
unit matrix.  For unitary pairing states the theory simplifies considerably
and the self-consistent equations for the gap function and the chemical
potential $\mu$, together with the expression for the ground state energy read
as
\begin{eqnarray}
\hat{\Delta}_{\bf k}&=&{1\over L}\sum_{\bf p} V_{{\bf k}{\bf p}}^{\rm sc}
{\hat\Delta}_{\bf p}
\tanh\left({E_{\bf p}\over 2T}\right),
\label{eq:sc_gap}
\\
N&=&\sum_{\bf k}\left[1-{\xi_{\bf k}\over E_{\bf k}}
\tanh\left({E_{\bf k}\over 2T}\right)\right],
\nonumber
\\
E_{\rm GS}&=&-\sum_{\bf k}
{(E_{\bf k}-\xi_{\bf k})^2\over 2E_{\bf k}}+\mu N,
\nonumber
\end{eqnarray} 
where $E_{\bf k}=(\xi_{\bf k}^2+|\Delta_{\bf k}|^2)^{1/2}$ is the BCS
quasiparticle energy, $|\Delta_{\bf k}|^2=\sum_{\nu=0}^3 |d^\nu_{\bf k}|^2$ is
the spectroscopic gap, $\varepsilon_{\bf k}$ is the bare electron dispersion,
and $\xi_{\bf k}=\varepsilon_{\bf k}-\mu$.

\section{Density wave channel}
Next we turn to the discussion of instabilities of the symmetric state with
respect to the formation of bound particle-hole pairs with a nonvanishing
total momentum ${\bf Q}$. The finite value of ${\bf Q}$ implies the presence
of spatial modulations in the ground state, hence the name density wave
channel.  For the sake of simplicity, we specialize to the case when $2{\bf
Q}$ is an inverse lattice vector. In particular, this is relevant for the
$t$-$t^\prime$ Hubbard model at the Van Hove density, when ${\bf
Q}=(\pi,\pi)$; this latter case will be treated later as an explicit numerical
example.

The key quantities describing the symmetry broken phase are the angle-resolved
particle-hole order-parameter fields $d_{\bf k}^0=2^{-1}\sum_\sigma\langle
c^\dagger_{{\bar{\bf k}}\sigma} c_{{\bf k}\sigma} \rangle$ and ${\vec d_{\bf
k}}=2^{-1}\sum_{\alpha\beta}\langle c^\dagger_{{\bar{\bf k}}\alpha}
{\vec\sigma_{\alpha\beta}}c_{{\bf k}\beta} \rangle$, describing pairs with
spin zero and spin one, respectively.  Here we have introduced the
abbreviation ${\bar{\bf k}}={\bf k}+{\bf Q}$ and ${\vec\sigma_{\alpha\beta}}$
are the Pauli matrices. Let us notice that $d_{\bar{\bf k}}^0=(d_{\bf
k}^0)^\ast$ and ${\vec d_{\bar{\bf k}}}=({\vec d_{\bf k}})^\ast$.

Next we ask the question about the effective interactions in the density wave
channel. In order to single out into $H_1$ those processes which scatter the 
particle-hole pairs with total momentum ${\bf Q}$, we have to choose
\begin{equation*}
\Delta_{1234}=\cases{1&for 
${\bf k}_3-{\bf k}_1={\bf Q}$ or ${\bf k}_3-{\bf k}_2={\bf Q}$\\
0&otherwise\\}
\end{equation*}
Note that this choice leads to a spin rotation-invariant theory, since it
satisfies the criterion Eq.~\ref{eq:spin_rotation_invariance}.

Now let us calculate the expectation value of the Hamiltonian ${\tilde H}$ in
a density wave state $|{\tilde\psi_{\rm dw}}\rangle$,
$E=\langle{\tilde\psi_{\rm dw}}|{\tilde H}|{\tilde\psi_{\rm dw}}\rangle$.
After a tedious but straightforward calculation we find the result
\begin{eqnarray}
E=\sum_{{\bf k}\sigma}\varepsilon_{\bf k}f_{{\bf k}\sigma}
-{1\over L}\sum_{{\bf k},{\bf p}}
\left[V_{{\bf k} {\bf p}}^{\rm cdw}
d_{\bf k}^0 d_{\bf p}^0+
V_{{\bf k} {\bf p}}^{\rm sdw}
{\vec d_{\bf k}}\cdot{\vec d_{\bf p}}\right]+E_{\rm FL} ,
\label{eq:dw_energy}
\end{eqnarray}
in complete analogy with Eq.~\ref{eq:sc_energy}. The coefficients in
front of the order parameter fields are the sought effective
interactions. They read as
\begin{eqnarray*}
V_{{\bf k} {\bf p}}^{\rm cdw}=&-U+{U^2\over 2}\left[
\chi_{\rm pp}^\prime({\bf k}+{\bf p},\varepsilon_{\bf k}+\varepsilon_{\bf p})
+\chi_{\rm pp}^\prime({\bf k}+{\bf p},\varepsilon_{\bar{\bf k}}
+\varepsilon_{\bar{\bf p}})\right]
\\
&-U^2\left[
\chi_{\rm ph}^\prime({\bf k}-{\bar{\bf p}},
\varepsilon_{\bf k}-\varepsilon_{\bar{\bf p}})
+\chi_{\rm ph}^\prime({\bar{\bf k}}-{\bf p},\varepsilon_{\bar{\bf k}}
-\varepsilon_{\bf p})\right],
\\
V_{{\bf k} {\bf p}}^{\rm sdw}=&U-{U^2\over 2}\left[
\chi_{\rm pp}^\prime({\bf k}+{\bf p},\varepsilon_{\bf k}+\varepsilon_{\bf p})
+\chi_{\rm pp}^\prime({\bf k}+{\bf p},\varepsilon_{\bar{\bf k}}
+\varepsilon_{\bar{\bf p}})\right],
\end{eqnarray*}
where $\chi_{\rm pp}^\prime({\bf q},\omega)$ is the real part of the
particle-particle susceptibility,
\begin{equation*}
\chi_{\rm pp}({\bf q},\omega)={1\over L}\sum_{\bf K}
{{1-f^0_{\bf K}-f^0_{\bf q-K}}\over
{\varepsilon_{\bf K}+\varepsilon_{\bf q-K}-\omega-i0}}.
\end{equation*}
The imaginary part again does not enter due to the choice of $\Delta_{1234}$
and the actual occupation numbers have been replaced by their noninteracting
values.  Note that unlike in the superconducting case, the effective
interactions are different in the singlet and triplet sector.

Let us notice that both density wave interactions are real and have the
following symmetries: $V_{{\bf k} {\bf p}}=V_{{\bf p} {\bf k}} =V_{{\bar{\bf
k}} {\bar{\bf p}}}$. In analogy to the BCS theory, it is useful to introduce
the gap functions $\Delta_{\bf k}^0=L^{-1}\sum_{\bf p} V_{{\bf k}{\bf p}}^{\rm
cdw} d_{\bf p}^0$ and ${\vec\Delta_{\bf k}}=L^{-1}\sum_{\bf p} V_{{\bf k}{\bf
p}}^{\rm sdw} {\vec d_{\bf p}}$.  One verifies easily that $\Delta_{\bar{\bf
k}}^0=(\Delta_{\bf k}^0)^\ast$ and ${\vec \Delta_{\bar{\bf k}}}=({\vec
\Delta_{\bf k}})^\ast$. Let us define the auxiliary quantities $\delta_{\bf
k}=(\varepsilon_{\bf k}-\varepsilon_{\bar{\bf k}})/2$ and $\omega_{\bf
k}=(\varepsilon_{\bf k}+\varepsilon_{\bar{\bf k}})/2$ with the symmetry
properties $\delta_{\bar{\bf k}}=-\delta_{\bf k}$ and $\omega_{\bar{\bf
k}}=\omega_{\bf k}$, in terms of which the mean field Hamiltonian can be
written in a compact form as
\begin{equation*}
H=\sum_{\bf k}^\prime 
({\hat c}_{\bf k})^\dagger N_{\bf k}{\hat c}_{\bf k} 
+\sum_{{\bf k}} \left(d^0_{\bf k}\Delta^0_{\bf k}+
{\vec d}_{\bf k}\cdot {\vec\Delta}_{\bf k}\right),
\end{equation*} 
where the prime restricts the summation to only within the magnetic zone and
$({\hat c}_{\bf k})^\dagger= (c^\dagger_{{\bf k}\uparrow},c^\dagger_{{\bf
k}\downarrow}, c^\dagger_{{\bar{\bf k}}\uparrow},c^\dagger_{{\bar{\bf
k}}\downarrow})$ is a 4-component vector.  Furthermore we have introduced a
$4\times 4$ matrix $N_{\bf k}=\omega_{\bf k}{\bf I}+M_{\bf k}$, where ${\bf
I}$ is a unit $4\times 4$ matrix and
\begin{eqnarray*}
M_{\bf k}&=&
\left(\begin{array}{cc}
\delta_{\bf k}{\bf 1}  & 
-(\Delta^0_{\bf k})^\ast{\bf 1}-{\vec\Delta}_{\bf k}^\ast\cdot{\vec\sigma}\\
-\Delta^0_{\bf k}{\bf 1}-{\vec\Delta}_{\bf k}\cdot{\vec\sigma} 
& -\delta_{\bf k}{\bf 1}
\end{array}\right),
\end{eqnarray*}
with ${\bf 1}$ denoting a unit 2$\times$ 2 matrix.  Let us define the vector
field ${\vec Q}_{\bf k}=\Delta_{\bf k}^0{\vec\Delta}_{\bf k}^\ast
+(\Delta_{\bf k}^0)^\ast{\vec\Delta}_{\bf k}-i{\vec\Delta}_{\bf
k}\times{\vec\Delta}_{\bf k}^\ast$.  In analogy to the discussion of triplet
superconductors, we will call states with ${\vec Q}_{\bf k}=0$ as unitary. A
major simplification is that for unitary states, $M_{\bf k}^2$ is proportional
to a unit matrix, $M_{\bf k}^2=E_{\bf k}^2{\bf I}$ with $E_{\bf
k}=\sqrt{\delta_{\bf k}^2+|\Delta_{\bf k}|^2}$, where we have defined the
spectroscopic gap $|\Delta_{\bf k}|^2=\sum_{\nu=0}^3|\Delta^\nu_{\bf k}|^2$.
One checks easily that the eigenvectors of $M_{\bf k}$ are $E_{\bf k}$ and
$-E_{\bf k}$, both of them being doubly degenerate.  In what follows we will
develop an explicitly spin rotation-invariant mean field theory for unitary
states, closely following the standard treatment of triplet superconductivity
\cite{Balian63,Anderson75}.

Let us assume that the unitary transformation ${\hat c}_{\bf k}=U_{\bf
k}{\hat\gamma}_{\bf k}$ from the bare electrons described by $({\hat c}_{\bf
k})^\dagger$ to the new quasiparticles described by $({\hat\gamma}_{\bf
k})^\dagger= (\gamma^\dagger_{{\bf k}\uparrow},\gamma^\dagger_{{\bf
k}\downarrow}, \gamma^\dagger_{{\bar{\bf
k}}\uparrow},\gamma^\dagger_{{\bar{\bf k}}\downarrow})$ brings $N_{\bf k}$ to
a diagonal form, ${\tilde N}_{\bf k}=U^\dagger_{\bf k} N_{\bf k} U_{\bf k}=
{\rm diag}(E_{{\bf k}2},E_{{\bf k}2}, E_{{\bf k}1},E_{{\bf k}1})$, where
$E_{{\bf k}1}=\omega_{\bf k}+E_{\bf k}$ and $E_{{\bf k}2}=\omega_{\bf
k}-E_{\bf k}$. Then the diagonalized Hamiltonian reads as
\begin{equation*}
H=\sum_{{\bf k}\sigma}^\prime 
\left[
E_{{\bf k}2}\gamma^\dagger_{{\bf k}\sigma}\gamma_{{\bf k}\sigma}+ 
E_{{\bf k}1}\gamma^\dagger_{{\bar{\bf k}}\sigma}\gamma_{{\bar{\bf k}}\sigma} 
\right]
+\sum_{{\bf k}}\left(d^0_{\bf k} \Delta^0_{\bf k}
+{\vec d}_{\bf k}\cdot {\vec\Delta}_{\bf k}\right),
\end{equation*}
and therefore $\langle {\hat \gamma}_{\bf k}({\hat \gamma}_{\bf
k})^\dagger\rangle= 2^{-1}\left[{\bf I} +\tanh\left({{\tilde N}_{\bf k}/
2T}\right)\right]$. Transforming back to the ${\hat c}_{\bf k}$ operators, we
thus have
\begin{equation}
\langle {\hat c}_{\bf k}({\hat c}_{\bf k})^\dagger\rangle=
U_{\bf k}\langle {\hat\gamma}_{\bf k}
({\hat\gamma}_{\bf k})^\dagger\rangle U_{\bf k}^\dagger=
{1\over 2}\left[{\bf I}+\tanh\left({N_{\bf k}\over 2T}\right)\right].
\label{eq:self_consistent_sdw}
\end{equation}
Expanding the function $\tanh x$ into the Taylor series, making use of the
binomial formula for powers of $N_{\bf k}=\omega_{\bf k}{\bf I}+M_{\bf k}$,
and taking into account that $M_{\bf k}^2=E_{\bf k}^2{\bf I}$, we find
\begin{eqnarray*}
\fl
\tanh\left({N_{\bf k}\over 2T}\right)
=
{{\bf I}\over 2}\left[\tanh\left({E_{{\bf k}1}\over 2T}\right)
+\tanh\left({E_{{\bf k}2}\over 2T}\right)\right]
+{M_{\bf k}\over 2E_{\bf k}}
\left[\tanh\left({E_{{\bf k}1}\over 2T}\right)
-\tanh\left({E_{{\bf k}2}\over 2T}\right)\right].
\end{eqnarray*}
Inserting this result to Eq.~\ref{eq:self_consistent_sdw} and comparing the
components of the matrices on both sides we find
\begin{eqnarray*}
n_{{\bf k}\sigma}&=&{1\over 2}
\left[f_{{\bf k}1}\left(1+{\delta_{\bf k}\over E_{\bf k}}\right)
+f_{{\bf k}2}\left(1-{\delta_{\bf k}\over E_{\bf k}}\right)
\right],
\\
n_{{\bar{\bf k}}\sigma}&=&{1\over 2}
\left[f_{{\bf k}1}\left(1-{\delta_{\bf k}\over E_{\bf k}}\right)
+f_{{\bf k}2}\left(1+{\delta_{\bf k}\over E_{\bf k}}\right)
\right],
\\
{\vec d}_{\bf k}&=&{\vec\Delta}_{\bf k}^\ast
{f_{{\bf k}2}- f_{{\bf k}1}\over E_{{\bf k}1}-E_{{\bf k}2}},
\\
d^0_{\bf k}&=&(\Delta^0_{\bf k})^\ast
{f_{{\bf k}2}- f_{{\bf k}1}\over E_{{\bf k}1}-E_{{\bf k}2}},
\end{eqnarray*}
and $\langle c^\dagger_{{\bf k}\sigma}c_{{\bf k}-\sigma}\rangle=0$.  In the
above equations we have used the notation $n_{{{\bf k}}\sigma}=\langle
c^\dagger_{{\bf k}\sigma}c_{{\bf k}\sigma}\rangle$ and $f_{{\bf k}i}=f(E_{{\bf
k}i}-\mu)$ where $f(x)$ is the Fermi-Dirac distribution function.  Note that
$f_{\bar{\bf k}i}=f_{{\bf k}i}$ since $E_{\bar{\bf k}i}=E_{{\bf k}i}$.  From
here we finally find the self-consistent equations for the gap matrix and for
the chemical potential, as well as the corresponding ground state energy:
\begin{eqnarray}
{\vec\Delta}_{\bf k}&=&{1\over L}\sum_{\bf p} V_{{\bf k}{\bf p}}^{\rm sdw}
{\vec\Delta}_{\bf p}^\ast
{f_{{\bf p}2}- f_{{\bf p}1}\over E_{{\bf p}1}-E_{{\bf p}2}},
\label{eq:sdw_gap}
\\
\Delta^0_{\bf k}&=&{1\over L}\sum_{\bf p} V_{{\bf k}{\bf p}}^{\rm cdw}
(\Delta^0_{\bf p})^\ast
{f_{{\bf p}2}- f_{{\bf p}1}\over E_{{\bf p}1}-E_{{\bf p}2}},
\label{eq:cdw_gap}
\\
N&=&\sum_{\bf k}[f_{{\bf k}1}+f_{{\bf k}2}],
\nonumber
\\
E_{\rm GS}&=&\sum_{\bf k}\left[E_{{\bf k}1}f_{{\bf k}1}
+E_{{\bf k}2}f_{{\bf k}2}
+|{\Delta}_{\bf k}|^2{f_{{\bf k}2}- f_{{\bf k}1}
\over E_{{\bf k}1}-E_{{\bf k}2}}\right].
\nonumber
\end{eqnarray}
Note that the summations in these equations run over the full Brillouin
zone. It should be stressed that the simple BCS-like form of
Eqs.~\ref{eq:sdw_gap},\ref{eq:cdw_gap} obtains only for unitary states with
${\vec Q}_{\bf k}=0$.

In what follows, we do not allow for the simultaneous presence of charge and
spin density wave order. Nevertheless, even with this simplification the
number of possible phases turns out to be quite large.  For instance, on a
square lattice there are ten different symmetry breaking patterns in the
singlet sector, since $\Delta^0_{\bf k}$ may transform according to one of the
five irreducible representations of the point group and there is an additional
double degeneracy associated with the parity of the gap function under the
translation in momentum space, ${\bf k}\rightarrow{\bar{\bf k}}$.  Namely, if
we decompose the order parameter to its real real and imaginary parts,
$\Delta^0_{\bf k}=x_{\bf k}+iy_{\bf k}$, the gap equation can be written as
\begin{eqnarray*}
\left(\begin{array}{l}
x_{\bf k}\\y_{\bf k}
\end{array}\right)
={1\over L}\sum_{\bf p} V_{{\bf k}{\bf p}}^{\rm cdw}
\left(\begin{array}{r}
x_{\bf p}\\-y_{\bf p}
\end{array}\right)
{f_{{\bf p}2}- f_{{\bf p}1}\over E_{{\bf p}1}-E_{{\bf p}2}}.
\end{eqnarray*}
As observed by Hankevych and Wegner \cite{Hankevych03}, the components with
the symmetry properties $x_{\bar{\bf k}}=x_{\bf k}$ and $y_{\bar{\bf
k}}=-y_{\bf k}$ are described by coupling constants with opposite signs.

As regards the triplet sector, based on the analogy of Eq.~\ref{eq:sdw_gap}
with the triplet superconductor case \cite{Balian63,Anderson75}, we expect
that for degenerate representations of the point group, energy is minimized by
Balian-Werthamer-like states involving complex patterns of ${\vec\Delta}_{\bf
k}$.  For square lattices only the $p$ representation is degenerate and since
that symmetry sector is not important for the model studied in Section 6, in
this paper we discuss only states of the type ${\vec\Delta}_{\bf k}=(x_{\bf
k}+iy_{\bf k}){\vec n}$, where ${\vec n}$ is a fixed direction in spin space
and $x_{{\bf k}},y_{{\bf k}}$ are real functions with the symmetry properties
$x_{\bar{\bf k}}=x_{\bf k}$ and $y_{\bar{\bf k}}=-y_{\bf k}$.  Repeating the
argument for the charge density waves, we find again that the components
$x_{{\bf k}}$ and $y_{\bf k}$ are described by coupling constants with
opposite signs.

\section{Landau channel}
In this Section we consider instabilities of the high temperature state with
respect to the formation of bound particle-hole pairs with total momentum
${\bf q}=0$.  In this so-called Landau channel we therefore choose
\begin{equation*}
\Delta_{1234}=\cases{1&for 
${\bf k}_3={\bf k}_1$ or ${\bf k}_3={\bf k}_2$\\
0&otherwise\\}
\end{equation*}
Let us introduce the angle resolved singlet and triplet particle-hole fields,
$n_{\bf k}=2^{-1}\sum_\sigma\langle c^\dagger_{{{\bf k}}\sigma} c_{{\bf
k}\sigma} \rangle$ and ${\vec d_{\bf k}}= 2^{-1}\sum_{\alpha\beta}\langle
c^\dagger_{{{\bf k}}\alpha} {\vec\sigma_{\alpha\beta}}c_{{\bf k}\beta}
\rangle$, respectively. In terms of $n_{\bf k}$ and ${\vec d_{\bf k}}$, the
expectation values of $H_1$ and ${\tilde H}_2$ read as
\begin{eqnarray*}
\langle H_1\rangle&=&{U\over L}\sum_{{\bf k}{\bf p}}
\left[n_{\bf k} n_{\bf p}-{\vec d}_{\bf k}\cdot{\vec d}_{\bf p}\right],
\\
\langle {\tilde H}_2\rangle&=&
{U^2\over L^2}\sum_{\{123\}}
{1-n_1-n_2
\over\varepsilon_3+\varepsilon_4-\varepsilon_1-\varepsilon_2}
\left[n_3 n_4-{\vec d}_3\cdot{\vec d}_4\right].
\end{eqnarray*}
A crucial difference between the singlet and triplet fields is as follows.
The triplet fields ${\vec d_{\bf k}}$ vanish in the symmetric high-temperature
phase and we expect that, in the weak coupling limit, the development of a
finite ${\vec d_{\bf k}}$ does not substantially change the electron
distribution function. Therefore we can construct the triplet Landau
interaction function as a second derivative of the interaction energy $\langle
H_1+{\tilde H}_2\rangle$ with respect to ${\vec d}_{\bf k}$. The derivatives
can be evaluated for undeformed Fermi surfaces, leading to the result
\begin{eqnarray*}
V^{\rm t}_{{\bf k}{\bf p}}&=&U-U^2\chi^\prime_{\rm pp}({\bf p}+{\bf k},
\varepsilon_{\bf p}+\varepsilon_{\bf k}).
\end{eqnarray*}

Contrary to the triplet case, the singlet field $n_{\bf k}$ does not vanish
even in the noninteracting system. If we expand the interaction energy
$\langle H_1+{\tilde H}_2\rangle$ with respect to deviations of $d_{\bf k}^0$
from its value in the noninteracting system, besides quadratic terms there
appear also linear terms in the deviation. Within the Landau Fermi liquid
theory, this means that the quasiparticle dispersion relation is modified with
respect to the bare spectrum. However, since in the discussion of
superconductivity and of the density waves we have not taken into account the
renormalization of the spectrum, in order to keep the same level of
approximation, we will consider only those singlet sector Landau instabilities
which occur in the non $s$-wave channel, i.e. we assume that $n_{\bf
k}=f^0_{\bf k}+d_{\bf k}^0$ where $f^0_{\bf k}$ is the noninteracting
distribution and $d_{\bf k}^0$ transforms with respect to a nontrivial
representation of the point group. With this restriction the terms linear in
$d_{\bf k}^0$ vanish and we can write
\begin{eqnarray}
E=\sum_{{\bf k}\sigma}\varepsilon_{\bf k}f_{{\bf k}\sigma}
-{1\over L}\sum_{{\bf k},{\bf p}}
\left[V_{{\bf k} {\bf p}}^{\rm s}
d_{\bf k}^0 d_{\bf p}^0+
V_{{\bf k} {\bf p}}^{\rm t}
{\vec d_{\bf k}}\cdot{\vec d_{\bf p}}\right]+E_{\rm FL},
\label{eq:landau_energy}
\end{eqnarray}
where
\begin{eqnarray*}
V^{\rm s}_{{\bf k}{\bf p}}= U^2\left[\chi^\prime_{\rm pp}
({\bf p}+{\bf k},\varepsilon_{\bf p}+\varepsilon_{\bf k})
-2\chi^\prime_{\rm ph}
({\bf p}-{\bf k},\varepsilon_{\bf p}-\varepsilon_{\bf k})\right].
\end{eqnarray*}
Note that both in the singlet and in the triplet sectors, the interaction
matrix is real and symmetric, $V_{{\bf k}{\bf p}}=V_{{\bf p}{\bf k}}$.  After
introducing the real gap functions $\Delta^0_{\bf k}=L^{-1} \sum_{\bf
p}V_{{\bf k}{\bf p}}^s d^0_{\bf p}$ and ${\vec\Delta}_{\bf k}=L^{-1} \sum_{\bf
p}V_{{\bf k}{\bf p}}^t{\vec d}_{\bf p}$, the mean field Hamiltonian can be
written as
\begin{equation*}
H=\sum_{\bf k}(\hat{c}_{\bf k})^\dagger 
\left[(\varepsilon_{\bf k}-\Delta^0_{\bf k}){\bf 1}
-{\vec\Delta}_{\bf k}\cdot\vec{\sigma}\right]\hat{c}_{\bf k} 
+\sum_{\bf k}\left[d^0_{\bf k}\Delta^0_{\bf k}
+{\vec d}_{\bf k}\cdot{\vec\Delta}_{\bf k}\right],
\end{equation*}
where $(\hat{c}_{\bf k})^\dagger= (c^\dagger_{{\bf k}\uparrow},c^\dagger_{{\bf
k}\downarrow})$.  Consider a unitary transformation $\hat{c}_{\bf k}=U_{\bf
k}\hat{\gamma}_{\bf k}$ to the quasiparticle operators $(\hat{\gamma}_{\bf
k})^\dagger= (\gamma^\dagger_{{\bf k}-},\gamma^\dagger_{{\bf k}+})$ with
\begin{eqnarray*}
U_{\bf k}={1\over\sqrt{2|{\vec\Delta}_{\bf k}|}}\left(\begin{array}{cc}
\sqrt{|{\vec\Delta}_{\bf k}|+\Delta^z_{\bf k}}, & -\sqrt{|{\vec\Delta}_{\bf
k}|-\Delta^z_{\bf k}} \\{\Delta^x_{\bf k}+i \Delta^y_{\bf k} \over
\sqrt{|{\vec\Delta}_{\bf k}|+\Delta^z_{\bf k}}}, & {\Delta^x_{\bf k}+i
\Delta^y_{\bf k} \over \sqrt{|{\vec\Delta}_{\bf k}|-\Delta^z_{\bf k}}}
\end{array}\right),
\end{eqnarray*}
where we have introduced $|{\vec\Delta}_{\bf k}|^2=\sum_{i=1}^3(\Delta_{\bf
k}^i)^2$.  One checks readily that the Hamiltonian is diagonal in the new
basis,
\begin{equation*}
H=\sum_{\bf k}\left[ E_{{\bf
k}-}\gamma^\dagger_{{\bf k}-}\gamma_{{\bf k}-} +E_{{\bf
k}+}\gamma^\dagger_{{\bf k}+}\gamma_{{\bf k}+} +d^0_{\bf k}\Delta^0_{\bf k}
+{\vec d}_{\bf k}\cdot{\vec\Delta}_{\bf k} \right],
\end{equation*} 
with energy eigenvalues $E_{{\bf k}+}=\varepsilon_{\bf k}-\Delta^0_{\bf
k}+|{\vec\Delta}_{\bf k}|$ and $E_{{\bf k}-}=\varepsilon_{\bf k}-\Delta^0_{\bf
k}-|{\vec\Delta}_{\bf k}|$.  Repeating the argument leading to
Eq.~\ref{eq:self_consistent_sdw} we find
\begin{eqnarray*}
\langle {\hat c}_{\bf k}({\hat c}_{\bf k})^\dagger\rangle=
U_{\bf k}
\left(\begin{array}{cc}
1-f_{{\bf k}-} & 0
\\0 & 1-f_{{\bf k}+}
\end{array}\right)
U_{\bf k}^\dagger.
\end{eqnarray*}
Making use of the formula $U_{\bf k}\sigma^zU_{\bf
k}^\dagger={\vec{\sigma}}\cdot{\vec\Delta}_{\bf k}/|{\vec\Delta}_{\bf k}|$, the
right hand side can be computed explicitly.  Comparing the matrix elements of
the left and right hand sides we find
\begin{eqnarray*}
{\vec d}_{\bf k}={\vec\Delta}_{\bf k}
{f_{{\bf k}-}-f_{{\bf k}+}\over E_{{\bf k}+}-E_{{\bf k}-}},
\qquad
d^0_{\bf k}={1\over 2}\left[f_{{\bf k}-}+f_{{\bf k}+}\right]-f^0_{\bf k}.
\end{eqnarray*}
From here follow the self-consistent equations for the gap functions and the
chemical potential, as well as the corresponding ground state energy:
\begin{eqnarray}
\fl
{\vec\Delta}_{\bf k}={1\over L}\sum_{\bf p} V^{\rm t}_{{\bf k}{\bf p}}
{\vec\Delta}_{\bf p}
{f_{{\bf p}-}-f_{{\bf p}+}\over E_{{\bf p}+}-E_{{\bf p}-}},
\label{eq:tlandau_gap}
\\
\fl
\Delta^0_{\bf k}={1\over L}\sum_{\bf p} V^{\rm s}_{{\bf k}{\bf p}}
\left[{1\over 2}(f_{{\bf p}-}+f_{{\bf p}+})-f^0_{\bf p} \right],
\label{eq:slandau_gap}
\\
\fl
N=\sum_{\bf k}\left[f_{{\bf k}-}+f_{{\bf k}+}\right],
\nonumber
\\
\fl
E_{\rm GS}=\sum_{\bf k}\left[
\left(E_{{\bf k}-}+{\Delta^0_{\bf k}\over 2}\right)f_{{\bf k}-}
+\left(E_{{\bf k}+}+{\Delta^0_{\bf k}\over 2}\right)f_{{\bf k}+}
+|{\vec\Delta}_{\bf k}|^2
{f_{{\bf k}-}-f_{{\bf k}+}\over E_{{\bf k}+}-E_{{\bf k}-}}\right].
\nonumber
\end{eqnarray}
We stress that Eqs.~\ref{eq:tlandau_gap},\ref{eq:slandau_gap} apply only if
singlet channel Pomeranchuk 'instabilities' are not allowed. In what follows,
we specialize to pure singlet or pure triplet instabilities.  Note that, as in
Section 4, for degenerate representations nontrivial states in the triplet
channel, in which ${\vec\Delta}_{\bf k}$ does not preserve the same direction
in spin space as one varies ${\bf k}$, are likely.

\section{Application: $t$-$t^\prime$ Hubbard model at the Van Hove density}

In the rest of this paper, we will study the phase diagram of the square
lattice $t$-$t^\prime$ Hubbard model in the plane with the ordinates
$t^\prime/t$ and electron filling, $\rho$. In order to be able to discuss also
the density wave states, we restrict ourselves to the so-called Van Hove line,
i.e. to those combinations of $t^\prime/t$ and $\rho$, which lead to a Fermi
surface of the noninteracting system which crosses the saddle points of the
dispersion at $(\pm\pi,0)$ and $(0,\pm\pi)$. With this choice the most
singular density wave susceptibility is expected at ${\bf Q}=(\pi,\pi)$,
satisfying the criterion that $2{\bf Q}$ is an inverse lattice vector.

All results reported in this Section were obtained on $L\times L$ lattices
with $L=128$ sites and periodic boundary conditions. The susceptibilities
entering the effective interactions have been written as convolutions and
calculated using the Fast Fourier Transform algorithm, as explained for the
particle-hole case in \cite{Zlatic00}. In the time direction, we have sampled
the interval $(0,\tau_{\rm max})$ with $N$ points. Thus the energy precision
is $\Delta\omega=2\pi/\tau_{\rm max}$ and the maximal frequency is
$\omega_{\rm max}=2\pi N/\tau_{\rm max}$.  In order that $\Delta\omega$ is
comparable to the energy precision in ${\bf k}$-space, we require
$\Delta\omega=8t/L$.  We also require $\omega_{\rm max}=64 t$ in order to
faithfully describe the high energy processes.  This leads us to the choice
$N=8L=1024$ time points and $\tau_{\rm max}=\pi L/(4t)=32\pi/t$.  In order to
reduce the numerical error, the time evolution was damped according to
$\exp(-\Gamma\tau)$ with $\Gamma=\Delta\omega/2=t/32$. The accuracy of the FFT
algorithm was checked by direct calculation of the susceptibilities. The
self-consistent equations were solved by the damped iterative method. The
condensation energy $E_{\rm cond}=E_{\rm norm}-E_{\rm GS}$ was calculated as
the difference between the energy of the symmetry broken state $E_{\rm GS}$
and the normal state energy $E_{\rm norm}=2\sum_{\bf k}\varepsilon_{\bf
k}f^0_{\bf k}$.  Note that stable states are described by a positive $E_{\rm
cond}$.

In what follows the symbols SC, CDW, SDW, SL, and TL stand for the
superconducting, charge density wave, spin density wave, singlet Landau
channel, and triplet Landau channel phases, respectively.  Spatial
representations of the point group of the square are denoted $s$, $d$,
$d_{xy}$, and $g$ (one-dimensional, even representations), and $p$
(two-dimensional, odd representation). Note that we have not introduced
separate symbols for singlet and triplet superconductors, since for
superconductors the spin sector is completely defined by the parity of the
spatial representation. The density wave instabilities are characterized by an
additional quantum number \cite{Hankevych03}, namely the parity of the order
parameter under translation in momentum space by ${\bf Q}$. Order parameters
satisfying $\Delta_{\bar{\bf k}}=\Delta_{\bf k}$ and $\Delta_{\bar{\bf
k}}=-\Delta_{\bf k}$ are distinguished by the suffix $1$ and $2$,
respectively.

\begin{figure}
\centerline{\includegraphics[width=6.5cm,angle=0]{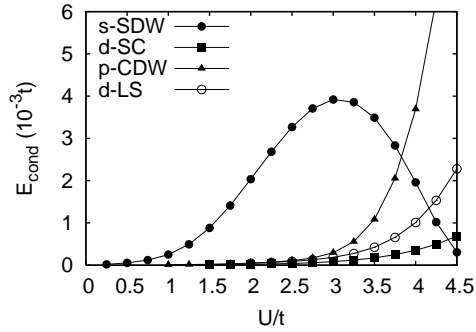}}
\caption{
\label{fig:tp0_00} 
Condensation energy as a function of $U$ at the Van Hove filling,
$t^\prime/t=0$, $\rho=1.0$, $L=128$. The $d$-SC state is degenerate with
$d$-SDW2 and $d$-CDW2 states.}
\end{figure}

In Fig.~\ref{fig:tp0_00} we plot the condensation energy as a function of $U$
for $t^\prime/t=0$ and at half filling, $\rho=1.0$. As expected, at weak
coupling the leading instability is towards an $s$-SDW1 (antiferromagnetic)
instability.  Around $U\approx 3t$ the condensation energy of the
antiferromagnetic state acquires a maximum.  At larger coupling constants
different symmetry breaking patterns dominate. Since the Hubbard model at half
filling and $t^\prime/t=0$ is expected to order antiferromagnetically for all
coupling constants \cite{Schrieffer89}, from Fig.~\ref{fig:tp0_00} we estimate
that our calculations are qualitatively correct up to $U\approx 3t$.

It should be noted that the overall shape of Fig.~\ref{fig:tp0_00}, and in
particular the prediction of the relative stability of the phases, is in
qualitative agreement with the calculation of the transition temperatures by
the flow equation method by Hankevych and Wegner \cite{Hankevych03}, see their
Fig.1. The only qualitative difference with respect to \cite{Hankevych03}
regards the relative stability of the $d$-LS ($d$-wave Pomeranchuk instability
\cite{Halboth00,Yamase00}) and $p$-CDW2 (band splitting \cite{Hankevych03})
phases: according to our calculation, the $p$-CDW2 phase is more stable than
the $d$-LS phase.  Hankevych and Wegner have also observed that for
$t^\prime/t=0$ and $\rho=1.0$, there is an additional symmetry which
guarantees that the $d$-SC state is degenerate with the $d$-SDW2 (triplet flux
phase \cite{Halperin68}) and $d$-CDW2 (singlet flux phase \cite{Halperin68})
states. This degeneracy can be proven also within our formalism and is nicely
satisfied by the data, thus providing a nontrivial check of the numerics.

\begin{figure}
\centerline{\includegraphics[width=6.0cm,angle=0]{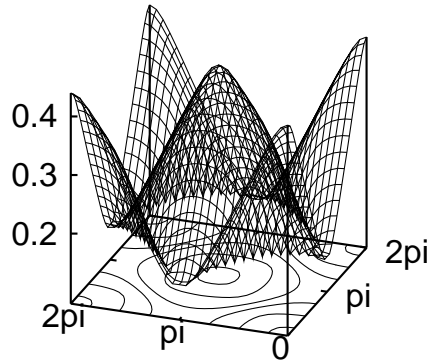}}
\caption{
\label{fig:gap_afm} 
Gap function in the antiferromagnetic ($s$-SDW1) phase. The parameters are the
same as in Fig.~\ref{fig:tp0_00} and $U=3t$.}
\end{figure}

Before proceeding it is worth pointing out that the gap function in the simple
antiferromagnetic ($s$-SDW1) phase is by no means featureless, as it would be
in a mean field theory.  Figure~\ref{fig:gap_afm} shows that $\Delta_{\bf k}$
is suppressed along the underlying square Fermi surface, the suppression being
largest close to the saddle points at $(\pm\pi,0)$ and $(0,\pm\pi)$. This
suppression is caused by the repulsive $U^2$ term in the effective interaction
$V_{{\bf k}{\bf p}}^{\rm sdw}$. Figure~\ref{fig:tp0_00} shows that the
repulsive term starts to dominate at $U>3t$. A proper discussion of the
spectroscopic implications of Fig.~\ref{fig:gap_afm} would require including
also the change of the normal state dispersion ($s$-SL channel), which we have
not attempted.

Summarizing the results at half filling, our data indicate that already at
$U\approx 3t$ where our approach should still apply, there exist nonnegligible
tendencies towards ordering in several nontrivial symmetry breaking states,
many of which have been extensively discussed in the context of the physics of
the cuprates: $d$-wave superconductivity, $d$-wave Pomeranchuk instability,
and the flux phases (in the singlet case also called $d$-density wave
phase \cite{Chakravarty01}). In what follows we discuss the evolution of these
tendencies as we move along the Van Hove line towards smaller electron
densities.

\begin{figure}
\centerline{\includegraphics[width=6.5cm,angle=0]{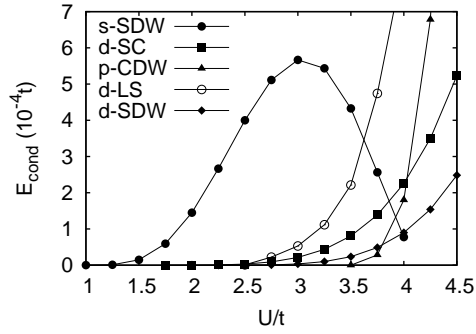}}
\caption{
\label{fig:tp0_05} 
Condensation energy as a function of $U$ at the Van Hove filling,
$t^\prime/t=0.05$, $\rho=0.959$, $L=128$.}
\end{figure}

\begin{figure}
\centerline{\includegraphics[width=6.5cm,angle=0]{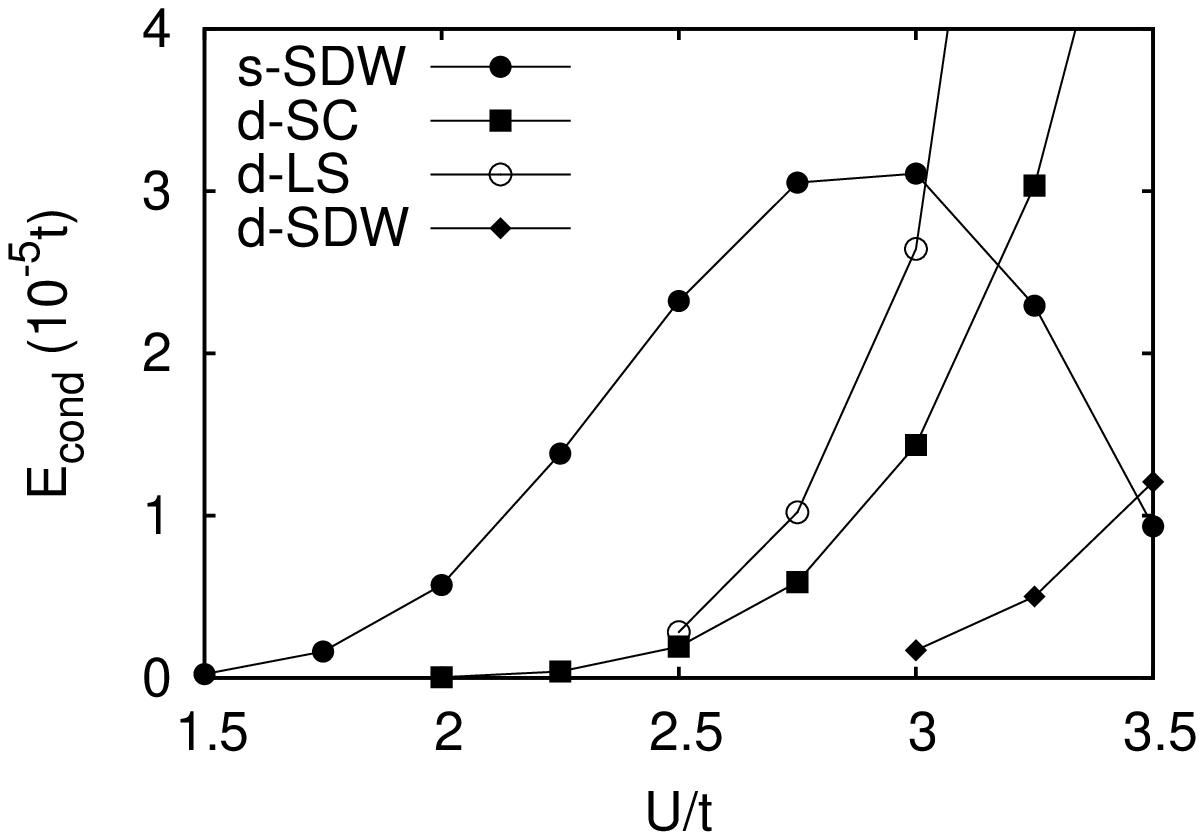}}
\caption{
\label{fig:tp0_10} 
Condensation energy as a function of $U$ at the Van Hove filling,
$t^\prime/t=0.10$, $\rho=0.918$, $L=128$.}
\end{figure}

\begin{figure}
\centerline{\includegraphics[width=6.5cm,angle=0]{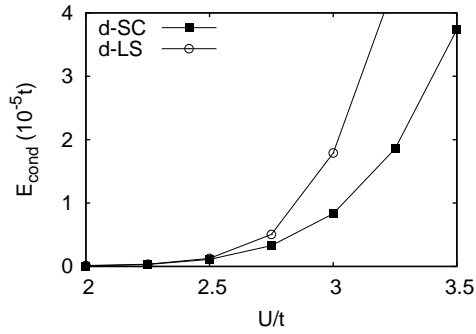}}
\caption{
\label{fig:tp0_15} 
Condensation energy as a function of $U$ at the Van Hove filling,
$t^\prime/t=0.15$, $\rho=0.875$, $L=128$.}
\end{figure}

Inspection of Figs.~\ref{fig:tp0_00} and \ref{fig:tp0_05}-\ref{fig:tp0_15}
shows that the antiferromagnetic ($s$-SDW1) condensation energy rapidly
diminishes with decreasing filling. This opens up the interesting possibility
that a nontrivial phase is stabilized at the Van Hove line.  The leading
subdominant instability at half filling is towards the band splitting phase
($p$-CDW2). Figures \ref{fig:tp0_00}, \ref{fig:tp0_05} and \ref{fig:tp0_10}
show that this phase quickly looses stability and never becomes dominant in
the region of applicability of our method. Moreover, the degeneracy between
the $d$-SC and the flux phases is lifted away from half filling: $E_{\rm
cond}(d{\rm SC})>E_{\rm cond}(d{\rm SDW2})>E_{\rm cond}(d{\rm CDW2})$.

Thus we are led to a study of the competition between the antiferromagnet, the
$d$-wave Pomeranchuk phase, and the $d$-wave superconductor. We find that,
along the singular Van Hove line, the Pomeranchuk instability is stronger than
the superconducting instability in the whole region we have studied, i.e. up
to $t^\prime/t=0.2$. Moreover, if we fix $U=3t$ and consider the competition
with the antiferromagnetic phase, we find that the $d$-wave Pomeranchuk phase
becomes stabilized for $t^\prime/t >(t^\prime/t)_c \approx 0.10$.

At the mean field level, the system at the Van Hove density is unstable
towards antiferromagnetism for infinitesimal coupling.  This raises the
following interesting question: Is it so that at sufficiently weak coupling,
when the mean field theory should apply, the system always has to order
antiferromagnetically?  Figures~\ref{fig:tp0_00}, \ref{fig:tp0_05}, and
\ref{fig:tp0_10} agree with this hypothesis. At smaller electron fillings we
cannot confirm it, since for our lattice size we cannot test reliably states
with small condensation energies.

\begin{figure}
\centerline{\includegraphics[width=6.5cm,angle=0]{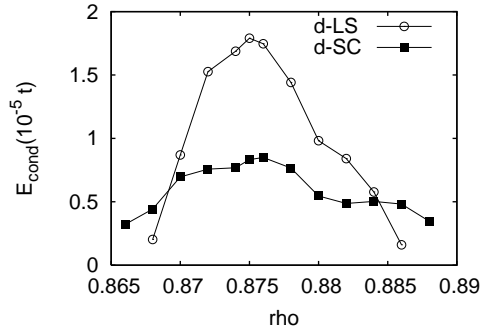}}
\caption{
\label{fig:density} 
Condensation energy as a function of $\rho$ in the vicinity of the Van Hove
filling, $t^\prime/t=0.15$, $U=3t$, $L=128$.}
\end{figure}

Our results indicate that, at the Van Hove filling, the particle-hole
instabilities are stronger than the superconducting ones.  In the rest of this
Section, we wish to study the relative stability of phases away from the Van
Hove filling.  First we notice that, in regions of the $t^\prime/t$ vs. $\rho$
plane lying close to the Van Hove line, the density wave instabilities are
expected to occur at wavevectors slightly different from ${\bf Q}=(\pi,\pi)$,
and the formalism developed in Section 4 does not apply. Therefore, in order
to work in a region where the density wave instabilities are safely
negligible, we have chosen to study the competition between the $d$-wave
Pomeranchuk phase and the $d$-wave superconductor in the vicinity of the Van
Hove point at $t^\prime/t=0.15$ for $U=3t$. The results are shown in
Fig.~\ref{fig:density}.  As expected, sufficiently far away from the Van Hove
filling (located at $\rho=0.875$), the superconducting phase is stabilized. An
unexpected feature of the data in Fig.~\ref{fig:density} is that the $d$-wave
Pomeranchuk phase is stable in an extremely narrow region around the Van Hove
line even at moderate coupling $U=3t$. This might explain why some of the
renormalization group studies observed the $d$-LS phase \cite{Halboth00} while
others did not \cite{Honerkamp01}.

\section{Conclusions}
In this work we have introduced a simple scheme which enables us to construct
effective Hamiltonians for correlated electron systems in a controlled
perturbative way. The method has been applied to the Hubbard model, but more
complicated models can be also studied.  As it stands, our scheme allows us to
treat superconductivity, Pomeranchuk instabilities (except for the singlet
instability transforming according to the trivial representation of the point
group), and density wave instabilities with ordering wavevector ${\bf Q}$
(subject to the restriction that $2{\bf Q}$ is an inverse lattice vector).
Both the singlet and the triplet sectors within each channel have been treated
and it has been observed that, for degenerate representations, the
phenomenology of the triplet particle-hole channels might be as rich as that
of the triplet superconductors \cite{Balian63,Anderson75}.

An effective implementation of the method, making use of the Fast Fourier
Transforms, has been developed. Lattices as large as $128\times 128$ are
tractable on PCs.  The method has been applied to the phase diagram of the
$t$-$t^\prime$ Hubbard model on the square lattice close to the Van Hove
filling, for $t^\prime/t$ in the interval between 0 and 0.2. Our results are
similar to those obtained by the flow equation method \cite{Hankevych03}: at
the Van Hove line, particle-hole instabilities are found to dominate. The
antiferromagnetic and the $d$-wave Pomeranchuk phases are stable for
$t^\prime/t<(t^\prime/t)_c$ and $t^\prime/t>(t^\prime/t)_c$, respectively. The
critical value of $(t^\prime/t)_c$ depends on the value of the interaction
strength $U$. The particle-hole instabilities are confined to an extremely
narrow stripe of the $t^\prime/t$ vs. $\rho$ plane around the Van Hove line
even at moderate coupling strength $U=3t$, leaving most of the phase space to
be dominated by the Kohn-Luttinger effect \cite{Kohn65}.

Straightforward future improvements of this work might include the study of
larger systems and a finite size scaling analysis of the results.  The theory
can be easily extended to finite temperatures.  It is also possible to take
into account the self-energy effects, or, in other words, the $s$-LS channel,
and also the change of $E_{\rm FL}$ in the symmetry breaking phases.
Furthermore, the method can be applied to density wave states with more
general wavevectors or to states with simultaneous presence of symmetry
breaking in different channels or symmetry sectors.

Turning to more speculative issues, an intriguing generalization would be to
consider condensates of more than two particles or holes. A bosonic condensate
would require a quadruplet of fields and an inspection of Eq.~\ref{eq:H_eff}
suggests that in order to find the effective interaction for such processes,
the perturbation expansion would have to be continued to higher orders.
Finite lifetime effects should be studied as well. This might require a
reorganization of perturbation theory in order to take into account the fact
that the energy conserving processes have been singled out into the tractable
term $H_1$.

A serious limitation of our scheme is its inability to treat states without
obvious order parameters, such as the Mott insulators. In such cases the
dynamical mean field theory (for a review, see \cite{Georges96}) seems to be
the method of choice. On the other hand, for problems requiring a high
resolution in momentum space, our method might be preferable.

\ack This work was supported by the Italian Ministry for Education,
Universities, and Research, by the Slovak Scientific Grant Agency under Grant
No.~VEGA-1/2011/05, and by the Centre of Excellence of the Slovak Academy of
Sciences CENG.


\end{document}